\documentclass[aps,prl,twocolumn,superscriptaddress]{revtex4-1}
\usepackage{graphicx}% Include figure files
\usepackage{dcolumn}% Align table columns on decimal point
\usepackage{bm}% bold math

\usepackage[normalem]{ ulem }
\usepackage{soul}
\usepackage{color}
\begin{document}

% Use the \preprint command to place your local institutional report
% number in the upper righthand corner of the title page in preprint mode.
% Multiple \preprint commands are allowed.
% Use the 'preprintnumbers' class option to override journal defaults
% to display numbers if necessary
%\preprint{}

%Title of paper
\title{Measuring Dirac Cones in a Sub-Wavelength Metamaterial}

\author{Simon Yves}
\affiliation{Institut Langevin, CNRS UMR 7587, ESPCI Paris, PSL Research University, 1 rue Jussieu, 75005 Paris, France}
\author{Thomas Berthelot}
\affiliation{CEA Saclay, IRAMIS, NIMBE, LICSEN, UMR 3685, F-91191, Gif sur Yvette, France}
\affiliation{KELENN Technology, 4 Avenue François Arago, 92160 Antony, France}
\author{Mathias Fink}
\affiliation{Institut Langevin, CNRS UMR 7587, ESPCI Paris, PSL Research University, 1 rue Jussieu, 75005 Paris, France}
\author{Geoffroy Lerosey}
\affiliation{Institut Langevin, CNRS UMR 7587, ESPCI Paris, PSL Research University, 1 rue Jussieu, 75005 Paris, France}
\affiliation{Greenerwave, ESPCI Paris Incubator PC’up, 6 rue Jean Calvin, 75005 Paris, France}
\author{Fabrice Lemoult}
\affiliation{Institut Langevin, CNRS UMR 7587, ESPCI Paris, PSL Research University, 1 rue Jussieu, 75005 Paris, France}
\email[]{fabrice.lemoult@espci.psl.eu}

\date{\today}% It is always \today, today,
             %  but any date may be explicitly specified

\begin{abstract}
The exciting discovery of bi-dimensional systems in condensed matter physics has triggered the search of their photonic analogues. In this letter, we describe a general scheme to reproduce some of the systems ruled by a tight-binding Hamiltonian in a locally resonant metamaterial: by specifically controlling the structure and the composition it is possible to engineer the band structure at will. We numerically and experimentally demonstrate this assertion in the microwave domain by reproducing the band structure of graphene, the most famous example of those 2D-systems, and by accurately extracting the Dirac cones. This is a direct evidence that opting for a crystalline description of those sub-wavelength scaled systems, as opposed to the usual description in terms of effective parameters, makes them a really convenient tabletop platform to investigate the tantalizing challenges that solid-state physics offer.
\end{abstract}

% PACS, the Physics and Astronomy
                             % Classification Scheme.
%\keywords{Suggested keywords}%Use showkeys class option if keyword
                              %display desired
\maketitle

Following the model of dielectric materials, it is possible to create artificial composite media which present at a macroscopic scale properties  that are not found in nature. Hence, these so-called \textit{metamaterials} are structured at scales that are much smaller than the wavelength of operation in the surrounding medium. Moreover, in the case of resonant subwavelength inclusions, they belong to a particular class of metamaterials, namely the locally resonant ones. Therefore, these systems are usually described by macroscopic effective properties, which can be designed at a mesoscopic scale in order to manipulate the wave propagation~\cite{Pendry1996,Engheta2006,Cai2010,Capolino2009}. Hence, several intriguing features have been demonstrated: negative refraction~\cite{Pendry2000,Smith2004}, cloaking~\cite{Schurig2006}, epsilon near-zero metamaterials~\cite{Engheta2013}, or unnaturally high refractive index~\cite{Choi2011} for example. We hereby focus in the microwave domain on a metamaterial which consists in a subwavelength lattice of metallic wires. This so-called wire medium has been first described by Pendry as a metallic mesostructure with a dilute number of electrons, hence allowing to tune the effective permittivity at low frequencies~\cite{Pendry1996}. Belov and his colleagues have also demonstrated that the uniaxial wire-medium, {\it ie.} the wires are all parallel, can be described by a non-local dispersive dielectric tensor~\cite{Belov2003}, which permitted to achieve subwavelength microwave imaging and sensing~\cite{Belov2005,Simovski2012}. 

Then, it has been demonstrated that a finite-sized version of such a medium supports below the light cone propagating modes that are known as spoof plasmon polaritons~\cite{Lemoult2010,Lemoult2011a}. In this latter case, one can easily understand the physics by adopting a microscopic point of view: all of the wires present the same length and behave as half-wavelength metallic resonators. The resonant interaction of each of them with the free space waves results in  an avoided crossing between the light line and the intrinsic resonance of the wires, known as a polariton. The dispersion relation presents two propagating branches, one of them corresponding to modes oscillating on a subwavelength scale~\cite{Lemoult2010,Lemoult2011a}, which are separated by a so-called hybridization band-gap. This generalizes to any medium made of subwavelength resonators and leads to the general class of locally resonant metamaterials also observed for example in acoustics~\cite{Lemoult2011b,Lanoy2015}, or with elastic waves~\cite{Rupin2014}.

\begin{figure}[htbt]
\includegraphics{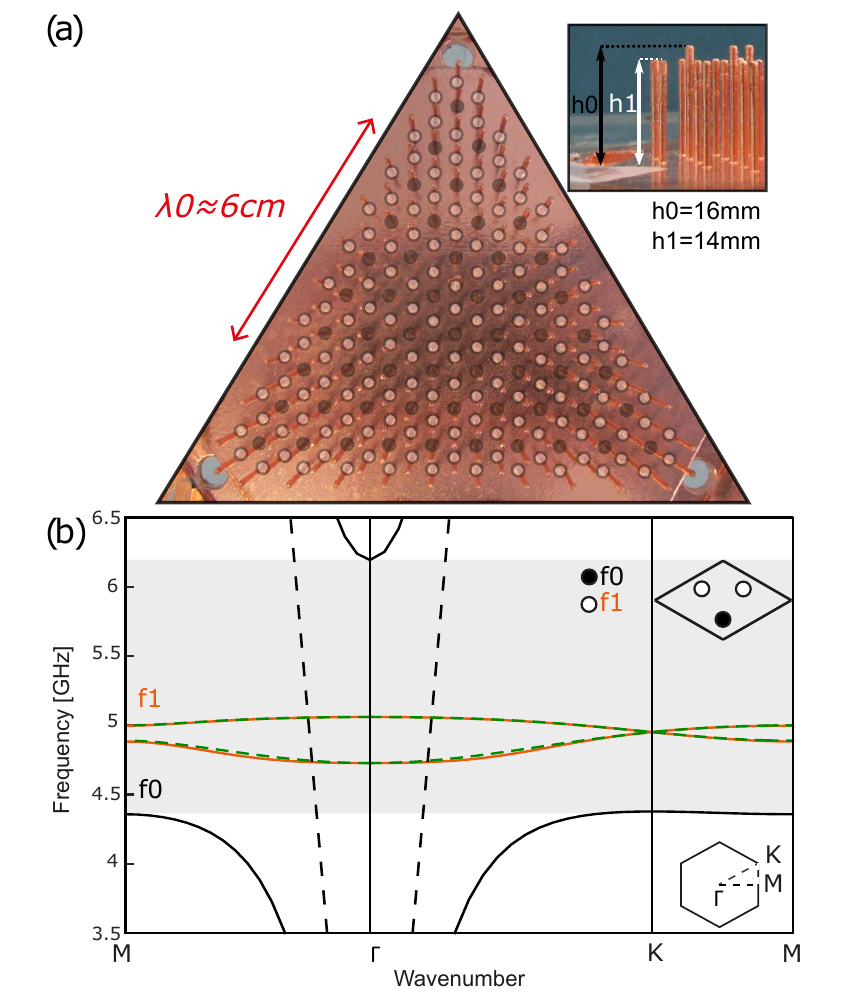}
	\caption{\label{Fig_1}(a) Photography of the sample (top view). It is composed of two sub-lattices of copper wires of different lengths $h_0$ (black) and  $h_1<h_0$ (white) which correspond to different resonance frequencies  $f_0$ and $f_1>f_0$. (b) Numerical dispersion relation of the system. Two bands (orange line) appear in the bandgap (shaded area) of the polariton due to the longer wires (black line). They cross at the $K$-point of the Brillouin zone, forming a Dirac cone far below the light cone (dotted line). }
\end{figure}

Describing the medium by its microscopic nature opens new possibilities. For example, modifying the unit cell, going from a simple resonant meta-atom to more complex meta-molecules, makes new features appear: a pair of resonators creates a negative refraction~\cite{Kaina2015}, or an hexamere of resonators can bring some topological effects~\cite{Yves2017}. This unit cell structuration is also present in photonic crystals~\cite{Joannopoulos2011}, albeit its consequences are visible at higher frequencies, when the wavelength scales with the lattice constant. Also, remembering the polaritonic bandgap, it is straightforward to locally insert a defect inside the medium: a detuned resonator with a resonance frequency falling within this bandgap~\cite{Lemoult2013,Kaina2013}. Because of the surrounding bandgap, waves are trapped on this defect site which acts as a subwavelength cavity. Adding a similar defect in the close vicinity of the first one enables the energy to â€œtunnelâ€ from one site to the other, allowing to guide waves on a subwavelength scale~\cite{Lemoult2013,Kaina2017}. Such an evanescent coupling between the defects is very similar to the nearest neighbor hopping term appearing in solid-state physics Hamiltonians. 
In this letter, we generalize this approach from a unidimensional line of defects to an infinite crystalline structure whose periodicity paves two dimensions of space. In this case, the bandgap of the polariton induces a nearest-neighbor coupling between the crystal's constituents.
We thus present a very genuine and straightforward recipe to emulate spinless condensed matter tight-binding Hamiltonians at a macroscopic scale within a locally resonant metamaterial. 

A well-known 2D example of such condensed-matter physics is the one of graphene, famous for the presence of Dirac cones within its band structure which condenses the peculiar electronic~\cite{Novoselov2005,Katsnelson2006,Geim2007,Beenakker2008,CastroNeto2009,Peres2010} and topological~\cite{Hasan2010} properties of the material.
Its Hamiltonian relies on two main features: the honeycomb arrangement of its carbon atoms and the nearest-neighbor hopping between them~\cite{Wallace1947,Reich2002}. Several macroscopic scale photonic analogues sharing these two characteristics have been proposed: Mie resonators placed on a ground plane~\cite{Kuhl2010,Bellec2013}, coupled cavities in photonic crystals~\cite{Fang2011}, or plasmonic nanostructures~\cite{Han2009}. Moreover macroscopic Dirac cones, whose origin lay in specific crystalline symmetries, also exist in photonic crystals~\cite{Ochiai2009,Bittner2010,Huang2011,He2015,Dong2016}. However these wavelength-scaled structured media are not comparable with the Hamiltonians we consider here as they are not ruled by  nearest-neighbor coupling.

In this study, we demonstrate that a locally resonant metamaterial is a useful platform to reproduce this interesting physics, but at a scale that is independant of the freespace wavelength.

We now focus on the electromagnetic locally resonant metamaterial made of wires acting as meta-atoms. They hereby consist of quarter-wavelength rods (1~mm diameter) on a ground plane and we designed the sample pictured in Fig.~\ref{Fig_1}.a by metallizing a 3D-printed ABS-based plastic medium, similarly to the procedure described in reference~\cite{Kaina2017}. The black disks correspond to a triangular sub-lattice of wires whose height is $h_0$=16~mm resonating at $f_0$, while the black circles correspond to a honeycomb sub-lattice of shorter wires of height $h_1$=14~mm thus resonating at a higher frequency $f_1$. The overall dimension of the sample is comparable to the freespace wavelength $\lambda_0$, demonstrating the deep subwavelength character of our approach. The polaritonic nature of the wave propagation within the triangular sub-lattice is responsible for the opening of a hybridization bandgap above the frequency $f_0$. This bandgap induces an evanescent coupling between the shorter wires, thus ensuring the nearest neighbor hopping, that is typical of graphene.

In order to check the relevance of the approach, we compute the corresponding numerical simulation which consists of solving the eigenfrequency problem in Comsol Multiphysics of a single unit cell with Bloch periodic boundary conditions for each wavenumber along the contour of the irreducible Brillouin zone. The dispersion relation along the main crystal directions (Fig.~\ref{Fig_1}.b) presents the two bands of the polariton (black lines), separated by a bandgap (shaded area) above the frequency $f_0$. Within the latter, two new propagating bands appear and they cross at the $K$-point of the first Brillouin zone, thus generating a Dirac cone. This crossing, relying on the biperiodicity of graphene's unit cell~\cite{Supplementary}, is well below the light cone (dotted line), meaning that it is of a subwavelength nature. They fit very well with the analytic dispersion relation of the graphene tight-binding Hamiltonian (dotted green line) and strongly confirms the pertinence of our method\footnote{The fitting second nearest-neighbor hopping term is $t_2=9.8~t_1$, $t_1$ being the nearest-neighbor hopping term}. 

	\begin{figure}
\includegraphics{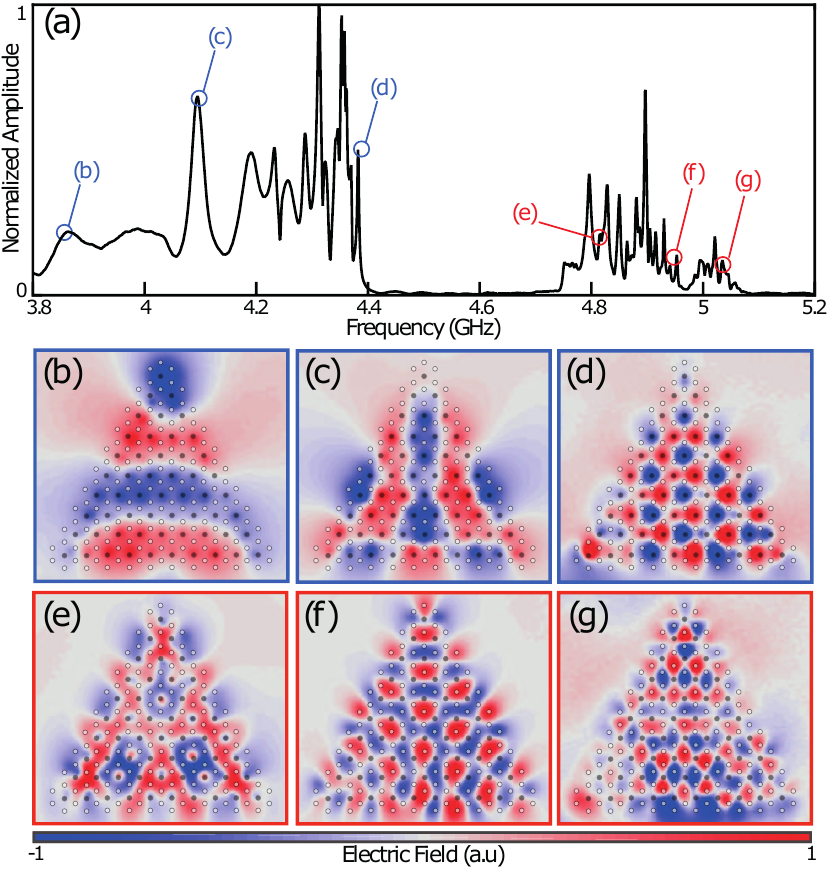}
	\caption{\label{Fig_2} (a) Experimental spectrum averaged on a large area within the bulk of the sample. Below are the experimental electric field maps (real part) corresponding to different frequencies in the polaritonic band (b,c,d)  and in the bands of defects (e,f,g).}
\end{figure}

\begin{figure*}
\includegraphics{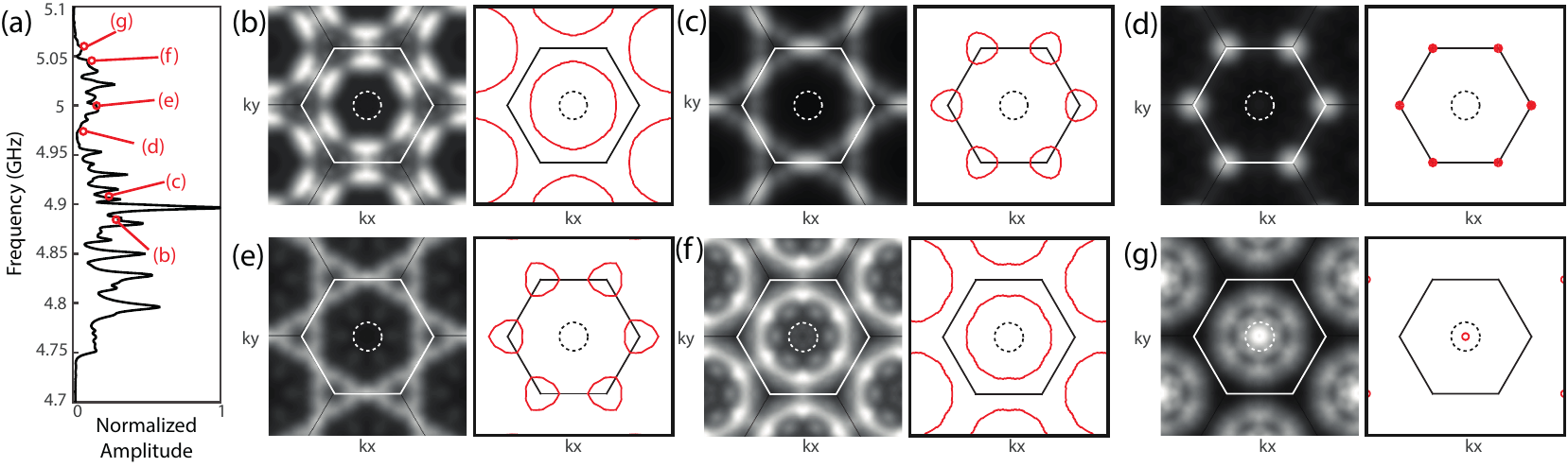}
	\caption{\label{Fig_3_horiz} (a) Zoom on the experimental spectrum averaged on a large area within the bulk of the sample. (b-g) On the left are the reconstructed reciprocal maps (see text for details) of some eigenmodes, and their corresponding extracted isofrequency contour on the right.}
\end{figure*}

To observe the subwavelength crystalline modes, a homemade antenna excites the sample in its close vicinity while a small probe, connected to a network analyzer, scans the electric near field right above the wires on the entire metamaterial. The frequency spectrum, averaged spatially over a large area in the center of the bulk (Fig.~\ref{Fig_2}.a) presents several resonance peaks: the source efficiently excitates bulk stationary modes within the finite-sized sample similarly to previous measurements~\cite{Lemoult2010}. Moreover, we clearly see two different sets of peaks, labeled in blue and red in Fig.~\ref{Fig_2}.a, which are separated by a bandgap. These bands appear at frequencies which well match with the numerical results presented previously.
As for the high frequency band of the polariton, it supports modes that are leaky in the free-space continuum and which cannot be measured in this near field experiment.

The experimental electric field maps corresponding to some of the previous peaks are shown in Fig.~\ref{Fig_2}.b-g. They clearly evidence the highly dispersive nature of the medium since the spatial scale observed in the maps varies considerably in a very short frequency range. Moreover, thanks to the small amount of dissipation within the copper wires, the modes, even the most subwavelength ones, are remarkably well-defined and distributed along the whole sample. Looking closely at the electric field maps of the first set of peaks (Fig.~\ref{Fig_2}.b-d), one can see that they are rather distributed on the triangular sub-lattice of longer wires (black disks). On the contrary, the experimental electric field maps corresponding to the second set of peaks (Fig.~\ref{Fig_2}.e-g) are essentially displayed on the honeycomb sub-lattice of defects ( black circles). Interestingly, a better look at the mode of Fig.~\ref{Fig_2}.e also indicates that the triangular lattice, responsible for the hybridization gap, is indeed $\pi$-shifted compared to the excitation of its neighbors. Nevertheless, this direct measurement of the sample's eigenmodes, although proving undoubtedly their subwavelength crystalline nature, is not yet sufficient to affirm that we have reproduced graphene's the band structure.

To evidence it, we now move to the reciprocal space in order to relate the frequency of the modes to their Bloch wavenumber. We concentrate on the second set of peaks measured in the spectral range from 4.75 to 5.1~GHz (Fig.~\ref{Fig_3_horiz}.a). In the reciprocal space, we know thanks to the Bloch theorem that the modes remain periodic and we can limit the study to the so-called first Brillouin zone, whose shape in this specific example is an hexagon. In order to extract the Bloch wavenumbers, we first carry out bi-dimensional spatial Fourier transforms of the measured modes. Our experimental protocol and the wave physics within the sample being linear, we are allowed to implement some transformations to these Fourier maps. We therefore sum the absolute values of the initial map with all of the maps obtained by applying one of the allowed symmetries. In our case, the symmetries corresponding to  those of the Bloch solution are: the $C_6$ rotational symmetry and the two mirror symmetries of the crystal, as well as the periodization of the first Brillouin zone imposed by the Bloch theorem. After using these symmetries and adequate summation of the normalized maps, we obtain the maps in the reciprocal space, centered on the first Brillouin zone (white hexagon), shown in Fig.~\ref{Fig_3_horiz}.b-g. 
From these we finally extract the points of maximum value so as to obtain the isofrequency contours. They are displayed on the right of the corresponding Fourier maps (red line), along with the freespace isofrequency contour (black-dashed line) to insist again on the fact that all happens on scales that are independent of the free-space wavelength. We notice very different behaviors depending on the frequency. Some of the contours (Fig.~\ref{Fig_3_horiz}.b,f and g) are circles centered at the $\Gamma$ point of the first Brillouin zone. The Bloch solution is almost isotropic and justifies the classical description in terms of effective index of refraction usually used in the context of metamaterials. On the contrary, some are strongly anisotropic which undoubtedly confirms the crystalline nature of the metamaterial under investigation. For example, Fig.~\ref{Fig_3_horiz}.c and e show contours that are centered around the K points in the corners of the first Brillouin zone. This is particularly interesting knowing that the Dirac cones of the honeycomb lattice are situated at these exact sites. Even more intriguing is the fact that these two contours are separated by the one of Fig.~\ref{Fig_3_horiz}.d which presents point singularities, again at these K points.

\begin{figure}
\includegraphics{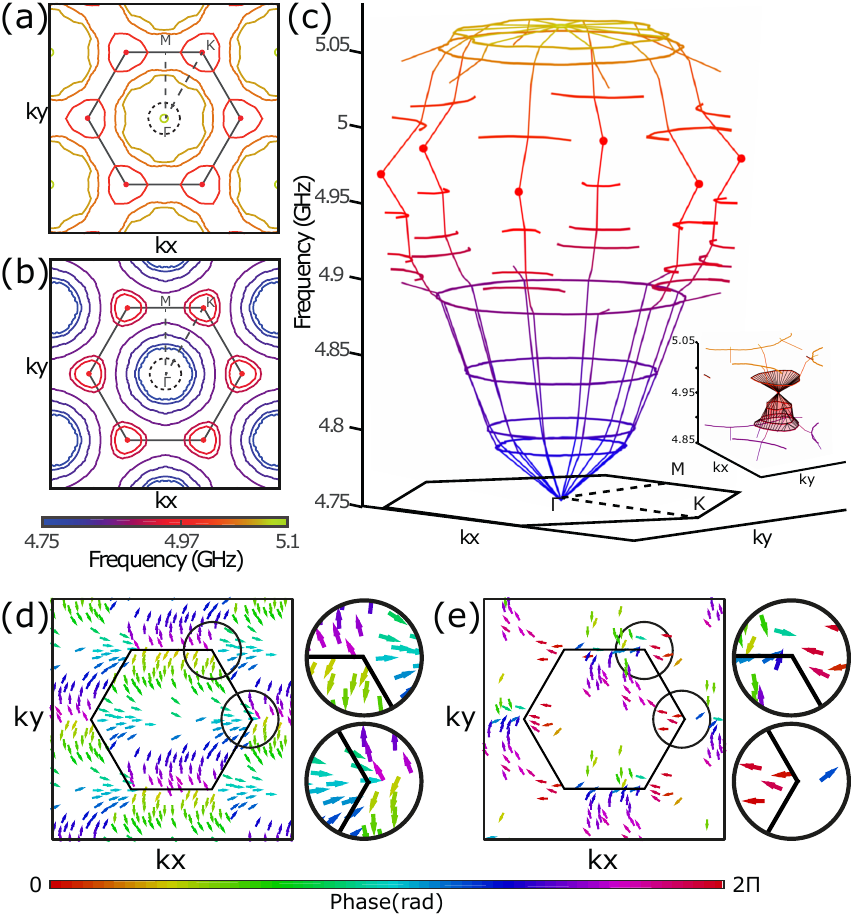}
	\caption{\label{Fig_4} Experimental subwavelength isofrequency contours of the upper (a) and lower (b) band of the honeycomb lattice of defects. (c) 3D representation of the total experimental band structure, which presents sub-wavelength Dirac cones (inset). (d)(resp. (e)) Experimental mapping of the phase shift between the two sites of the honeycomb unit cell for the low (resp. high) frequency band. Different vortices appear at the corners of the Brillouin zone.}
	\label{Fig4}	
\end{figure}

We eventually carry out the same procedure for a larger number of frequencies in order to reconstruct the band structure around these point singularities. We stack all of the obtained isofrequency contours and color them according to their frequency. It allows us to distinguish the two different bands of the dispersion relation. The lowest one, represented in Fig.~\ref{Fig_4}.b, is composed of concentric circles centered on the $\Gamma$-point, the center of the Brillouin zone. Their diameter increases along with their frequency showing the positive slope of the band. Then, they turn into circles centered at the $K$-points at the corners of the Brillouin zone, which demonstrates a gradually increasing anisotropy of the crystal. Their radius diminishes until these circles become single points situated at the six corners of the Brillouin zone. As for the remaining contours of the highest band, pictured in Fig.~\ref{Fig_4}.a, they present the inverse behavior compared to the lowest one. Indeed, they continuously move from the single point degeneracies at the K point to isotropic circles centered on the $\Gamma$ point of the Brillouin zone. This is related to a negative slope coherently to the dipolar nature of the modes of this branch~\cite{Kaina2015}.

A 3D representation of this experimental band structure is displayed on Fig.~\ref{Fig_4}.c. Lines linking the neighboring contours are added to enhance the perspective. This representation, along with Fig.~\ref{Fig_4}.a and b, demonstrates unambiguously the presence of Dirac cones at the $K$-points (inset).  We also insist on the fact that they are of very sub-wavelength nature, {\it ie.} they are far from the free space circle. Eventually, we stress that it has been very easy to span the cones since in our case the frequency plays the role of the Fermi level for real crystal. 
Moreover, these Dirac cones are known to exhibit peculiar behaviors related to the phase of the projection of the Dirac Hamiltonian onto the Pauli basis~\cite{CastroNeto2009}. The latter is directly obtained for one mode, corresponding to one wavenumber, by taking the phase shift between the two sites of the unit cell of the honeycomb lattice of defects. We then carry out this procedure for both of the two defect bands, by averaging the phase shift measurement on the whole honeycomb sub-lattice. The experimental results are presented on Fig.~\ref{Fig_4}.d and e, for the lower and upper band respectively. They undeniably exhibit vortices at the corners of the Brillouin zone, whose features depend on the corner (inset), and are inverted between the two bands. Notably, these experimental maps are in a quite good agreement with their numerical counterparts~\cite{Supplementary}. These peculiar windings of the phase around the Dirac cones are closely related to their topological properties, namely quantized Berry phases equal to $\pm\pi$ according to the type of vortex.
As a summary, the strong versatility and great simplicity of our system allows to bring Dirac's physics at a subwavelength scale and to envision numerous intriguing solid-state physics scenarii such as compact topological properties for example.

As a conclusion, this experimental work shows that going beyond the usual effective parameter description of the locally resonant metamaterials permits to genuinely design a macroscopic analogue of graphene in the microwave regime. We have experimentally observed the famous Dirac cones and their corresponding Berry phases, as well as a remarkably clear transition from isotropic isofrequency contours centered around $\Gamma$ to two touching cones centered at the $K$-points. This procedure does not limit itself to the simple case of graphene, but generalizes to any tight-binding governed Hamiltonians. We believe that it makes locally resonant metamaterials a promising tabletop platform for the study of tantalizing condensed matter physics phenomena.

% If you have acknowledgments, this puts in the proper section head.
\begin{acknowledgments}
This work is supported by LABEX WIFI (Laboratory of Excellence within the French Program “Investments for the Future” ) under references ANR-10-LABX-24 and ANR-10-IDEX-0001-02 PSL*, and by Agence Nationale de la Recherche under reference ANR-16-CE31-0015.
\end{acknowledgments}

% Create the reference section using BibTeX:
%

\end{document}